\documentclass[aps,prl,preprint,superscriptaddress]{revtex4}
\usepackage{graphicx} 
\newcommand{\chipp}{\ensuremath{\chi^{\prime\prime}(\vec{q}, \omega)}}

\newcommand{\chpIa}{\ensuremath{\chi_I^{\prime\prime}(\vec{q}, \omega)_a}}
\newcommand{\chpD}{\ensuremath{\chi_P^{\prime\prime}(\vec{q}, \omega)}}
\newcommand{\tef}{\ensuremath{\tau_{\scriptstyle eff}}}
\newcommand{\tei}{\ensuremath{\tau_{\scriptstyle eI}}}
\newcommand{\ted}{\ensuremath{\tau_{\scriptstyle eP}}}
\newcommand{\cu}{$^{63}Cu$}
\newcommand{\ox}{$^{17}O$}
\begin{document}

\title{A pseudogap term in the magnetic response of the cuprate superconductors}
\author{R. E. Walstedt}
\email[]{walstedt@umich.edu}
\affiliation{Department of Physics, University of Michigan, Ann Arbor, MI 48106}
\author{T. E. Mason}
\affiliation{Oak Ridge National Laboratory, Oak Ridge, TN 37831}
\author{G. Aeppli}
\affiliation{London Centre for Nanotechnology and Department of Physics and Astronomy, UCL, London WC1E 6BS, UK}
\author{S. M. Hayden}
\affiliation{Department of Physics, University of Bristol, Bristol BS8 1TL, UK}
\author{H. A. Mook}
\affiliation{Oak Ridge National Laboratory, Oak Ridge, TN 37831}
\begin{abstract}
 
We combine neutron scattering (INS) data and NMR/NQR nuclear spin lattice relaxation rate (1/$T_1$) data to deduce the existence of a new contribution to the magnetic response \chipp\ in cuprate superconductors. This contribution, which has yet to be observed with INS, is shown to embody the magnetic pseudogap effects. As such, it explains the long-standing puzzle of pseudogap effects missing from cuprate INS data, dominated by stripe fluctuations, for \chipp\ at low energies.  For $La_{1.86}Sr_{0.14}CuO_4$ and $YBa_2Cu_3O_{6.5}$, the new term is the chief contributor to $1/T_1$ for $T\gg T_c$.
\end{abstract}

\maketitle

Cuprate superconductors are well-known for their unusual normal metallic state properties. Prominent among these is an extended 'pseudogap' regime located above the dome of superconducting transition temperatures $T_c$ (see e.g. a typical phase diagram \cite{Kivel2}). 
Theorists have attempted to connect features in the normal state phase diagram with the phenomenon of high temperature superconductivity itself. There are three main approaches. The first focuses on the disappearance of the Mott-Hubbard antiferromagnet, which is transformed into a valence bond state where mobile holes are naturally paired \cite{PWAPAL}.  The second concentrates on 'stripe' correlations as providing the environment needed for superconductivity \cite{Kivel}. The third hypothesizes that the cross-over into the pseudogap regime is actually a phase transition to a state with concealed long range order \cite{AjiV}.  Experimentally, we have been able to use inelastic neutron scattering (INS) as a function of temperature, composition and magnetic field to map stripe order and fluctuations \cite{Tran,AeppliMason,Stock}. 
In the joint NMR and INS analysis presented here we identify a novel low-frequency signal which is correlated with pseudogap formation.  

Beginning as an NMR effect \cite{WWRW,Alloul,GotoYas}, the pseudogap has been observed as a genuine charge-energy gap \cite{ARPES,Loes,Ding}, with excitations known as 'arc fermions', characterized in detail through recent ARPES studies \cite{UIC1,UIC2}.  Thermally induced changes of the Fermi surface, with concomitant behavior of arc fermion excitations, are clearly related to magnetic manifestations of the pseudogap.  
However, INS data for \chipp\ show only indirect manifestations of the pseudogap \cite{PCD,Hayden}.  
Meanwhile, systems such as La$_{1.86}$Sr$_{0.14}$CuO$_4$ (LSCO) \cite{AeppliMason} and YBa$_2$Cu$_3$O$_{6.5}$ (YBCO6.5) \cite{Stock} yield data for \chipp\ that consists at low frequencies of incommensurate, antiferromagnetically correlated peaks whose intensity exhibits $\omega/T$ scaling \cite{VLSR} from T$\sim$ 60K up to room temperature. Pseudogap effects are totally absent from such data.  Moreover, nuclear spin-lattice relaxation rates (1/$T_1$) for these systems \cite{Imai,WSC,Taki} are inconsistent with extrapolation of the INS results to NMR frequencies. 

      In this Letter we present a new, joint analysis of INS and NMR ($T_1$) data for the systems mentioned above, in which we deduce the existence of a pseudogap fluctuation term \chpD, which has not been identified by INS up to now. 
Thus, we write \chipp\ = \chpIa\ + \chpD, where \chpIa\ is the INS-measured term with incommensurate peaks along the $a$ axis. 
The term \chpD\ introduced here, which is nonzero in the fluctuating stripe phase, will be modeled below to interpret the $T_1$ data. 
Thus, not only does the strongly evidenced occurrence of such a term clearly explain the hitherto baffling omission of a pseudogap effect from data for \chipp\cite{AeppliMason,Stock}, it also accounts for the disparate behavior of $T_1$ for the planar \cu\ and \ox\ nuclear spins in these systems \cite{WSC,Taki}. We also show that the thermal and q-space behavior of \chpD\ is such that it could have easily been missed up to now by INS experiments on these systems. In sum, the present analysis addresses a major deficiency in our understanding of the anomalous normal-state physics of cuprates and will be of wide interest to theorists and experimentalists alike.

For the nuclear relaxation analysis we employ the formulation of the $T_1$ process pioneered by Uldry and Meier \cite{UM}, in which the relaxation rates are written
\begin{eqnarray}
    \frac{1}{^{63}T_{1c}}& =& \frac{\gamma_{63}^2}{2}\left[ A_{ab}^2 + 4B^2 + 2B^2(4K_{2} + K_{3a} + K_{3b})\right. \nonumber \\
   & & \left. \mbox{} + 4A_{ab}B(K_{1a} + K_{1b})\right]\tef
\label{63T1ef}
\end{eqnarray}
\begin{equation}
		\frac{1}{^{17}T_{1c}} =  \frac{\gamma_{17}^2}{4}[C_a^2 + C_b^2](2 + K_{1a} + K_{1b})\tef
\label{17T1ep}
\end{equation}
for \cu\ and \ox, the two nuclear species of interest. In these equations $A_{ab}$, $B$, $C_a$ and $C_b$ are hyperfine tensor components in units of 
Gauss per unit of spin \cite{WSC}.  The quantities $K_{n}$ represent the normalized dynamical spin-spin correlation functions \cite{UM,MilaRice,BSS}.  
Thus, $K_{n}$ = $4\langle\vec{S}_i\cdot\vec{S}_j\rangle$, where $\vec{S}_i$ and $\vec{S}_j$ are $n^{th}$ neighbor spins. 
The additional subscript $a,b$ indicates, for $n$ = 1 and 3, that the bond axis (i.e. $\vec{r}_{ij}$) lies along the $a$ or $b$ crystalline axis.
Using the fluctuation-dissipation formulation of $T_1$ \cite{Toru,MMP}, we express the $K_n$'s in terms of $\chipp_a$, taking $a$ as the discommensuration axis.  
The $K_n$'s (n = 1,2,3) for the first three neighbor pairs in the CuO$_2$ plane may then be written
\begin{equation}
K_{na,b}  = \frac{\int_nd\vec{q}\, g_{na,b}(\vec{q})[\chipp_a/\omega]_{\omega\rightarrow 0}}
{\int_nd\vec{q}\,[\chipp_a/\omega]_{\omega\rightarrow 0}},
\label{Kn}
\end{equation}
where $g_{1a,b}$ = $cos(q_{a,b}a)$; $g_2$ = $cos(q_aa)cos(q_ba)$; and $g_{3a,b}$ = $cos(2q_{a,b}a)$ in an obvious notation. Note that $K_2$ is independent of the discommensuration axis.  It is clear that $|K_n|\le$ 1 . 
Eq.(\ref{63T1ef}) and (\ref{17T1ep}) also employ the key parameter 
\begin{equation}
		 \tau_{eff}(T) = \frac{k_BT}{\mu_B^2}\int_nd\vec{q}
		 \left[\frac{\chi^{\prime\prime}(\vec{q},\omega )_a}{\omega}\right]_{\omega\rightarrow 0}, 
\label{Tauef}
\end{equation} 			
proportional to the 'local susceptibility', which acts as a correlation time that includes the particle statistics of the relevant carriers. \tef\ may also be estimated directly from INS data.  
Thus, we shall proceed by comparing the latter values of \tef\ with those obtained from $T_1$ data, using reasonable estimates of the other parameters in Eqs.(\ref{63T1ef}) and (\ref{17T1ep}).

\begin{figure}[htbp] 
\center
\includegraphics[bb= 0.4in 0in 2.9in 5in, scale = 0.9]{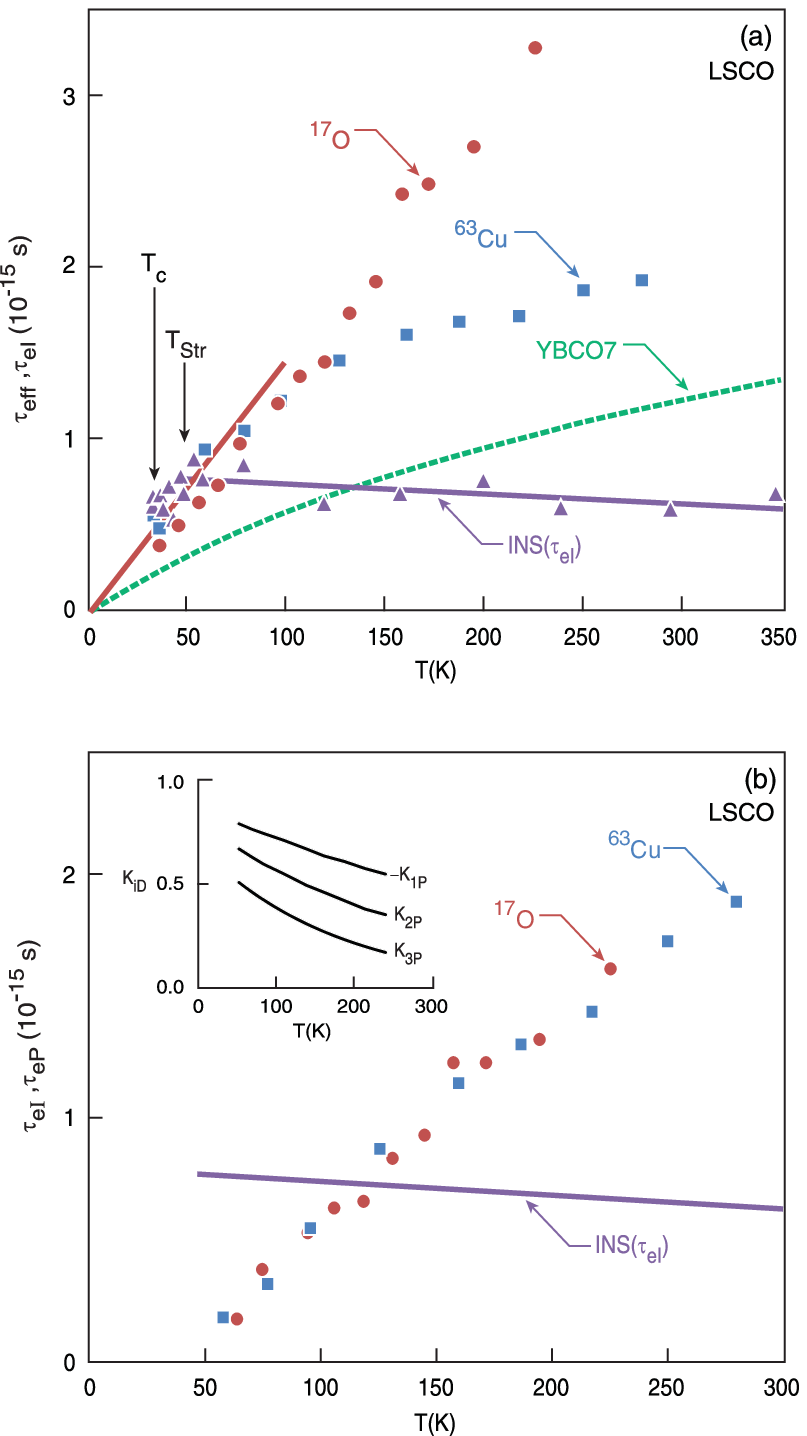}
\caption{(a) Values of $\tau_{eI}(T)$ and \tef\ determined in various ways for LSCO are plotted as functions of temperature.  Values of $\tau_{eI}(T)$ (solid triangles) were calculated with Eq.(4).  The solid line is a linear regression showing approximate $\omega/T$ scaling for $T > T_{Str} \sim$ 50K.  Values of \tef\ obtained from $T_1$ data with Eq.(1) and (2) are also plotted for \cu\ (squares) and \ox (circles), respectively.  The Korringa-like behavior of the $\tau$'s below $T_{Str}$ is highlighted by the solid red line. Data for \ox\ are scaled to that line for the analysis in part (b). For a general comparison, values of $\tef(T)$ for YBCO7 obtained by Uldry and Meier \cite{UM} for that system are shown as a dashed line.  (b) Values of $\tau_{eP}(T)$ obtained with Eqs.(5) and (6) using the same $T_1$ data as above are plotted against temperature. For this purpose, values of the $K_{nP}$'s derived from a squared-Lorentzian model are used.  The $K_{nP}$'s with their fitted temperature dependence are shown in the inset.  The solid line representing the behavior of $\tau_{eI}$ from part (a) is replotted here for comparison. See text for further discussion.
} 
\end{figure}

We begin by applying these equations to LSCO, using INS data for $\chipp_a$, designated \chpIa, in Eq.(\ref{Tauef}) to estimate \tef, 
which we denote \tei.
The resulting values \cite{Error}, plotted as solid triangles in Fig.1(a), exhibit approximate $\omega/T$ scaling (i.e. \tef\ = constant) above  the stripe onset temperature $T_{Str}\sim 50K$ \cite{VLSR}, in accord with the fluctuating stripe domains model of Zaanen et al. \cite{JZ1, JZ2}.   
Below $T_{Str}$, \tei(T) drops essentially linearly toward zero.  We note that $T_c\sim$ 35K for this sample \cite{AeppliMason}. 

 

The latter results are to be compared with estimates of \tef(T) derived from $T_1$ data \cite{Imai,WSC} using Eq.(1) and (2).   
To do this, we employ values of the HF constants derived from shift data \cite{HFConst} and values of the $K_{na,b}$'s calculated with Eq.(\ref{Kn}),
using a two-peak form factor fitted to INS data \cite{AeppliMason}.   
The $K_{nIa,b}$'s so determined vary only gradually with temperature.  As an example we mention values at T = 100K: $K_{1Ia}$ = -0.75; $K_{1Ib}$ = -0.97; $K_{2I}$ = 0.73; $K_{3Ia}$ = 0.16; $K_{3Ib}$ = 0.90.  One notes a sharp distinction between the $a$ and $b$-axis values as expected. 
For the \ox, this could result in the prediction of two widely different values of $1/{^{17}T_{1c}}$ for oxygen sites with Cu-O bonds aligned with 
the $a$ and $b$ axes, respectively.  
Since only a single rate was observed \cite{WSC}, it is presumed either that a flip-flop mechanism is present to maintain a single \ox\ nuclear spin temperature, or the stripe domain boundaries are fluctuating, so that each site automatically averages the two rates to yield a composite rate (given by Eq.(\ref{17T1ep})).    
Results so obtained are plotted in Fig.\,1(a) as blue squares and red circles for the $^{63}$Cu and $^{17}$O nuclear spins, respectively. For comparison, the value of \tef\ deduced for YBCO7 by Uldry and Meier \cite{UM} is shown as a dashed line, reflecting the fact that the $T_1$ process in LSCO is substantially stronger than that for YBCO7.  

Regarding the estimated values of \tef\ for LSCO in Fig.\,1(a), note that there is approximate Korringa-like behavior for all three data sets below $T$ = $T_{Str}$, where the agreement is good considering that there are no adjustable parameters.  The slight disparity in magnitudes is attributed to HF constant errors or differences in the widths of incommensurate peaks (i.e., of the $K_n$'s) among samples (see the YBCO6.5 case below).
Since the curves for \tef\ rise high above those for \tei\ at T $> T_{Str}$, the $T_1$ data give clear evidence for an additional term in \chipp\ as stated above.  We emphasize that there is no other realistic possibility among the well-documented $T_1$ mechanisms in solids.
Also noteworthy is the drastic difference between the \tef\ curves derived from the $^{63}$Cu and $^{17}$O $T_1$ data.  Values of \tef\ for these two measurements must actually be the same, so that the correlation properties of the new term \chpD\ are evidently rather different from those of \chpIa. 
  
Our next step is to take explicit account of the two terms in \chipp\ and rewrite Eqs.(1) and (2) as
\begin{eqnarray}
 \lefteqn{ \frac{1}{^{63}T_{1c}} = \frac{1}{^{63}T_{1Ic}}} \\
   & &+\, \frac{\gamma_{63}^2}{2}\left[ A_{ab}^2 + 4B^2(1 + 2K_{2P} + K_{3P}) + 8A_{ab}BK_{1P}\right]\ted \nonumber 
\label{63T1e2}
\end{eqnarray}
and
\begin{equation}
 \frac{1}{^{17}T_{1c}} =  \frac{1}{^{17}T_{1Ic}} + \frac{\gamma_{17}^2}{2}[C_a^2 + C_b^2]((1+K_{1P})\ted,
\label{17T1e2}
\end{equation}
where $K_{nP}$ and \ted\ are defined using \chpD\ in Eqs.(3) and (4), respectively.
In Eqs.(5) and (6) the $T_{1Ic}$'s are calculated with Eqs.(1) and (2), respectively, using \tef\ = \tei, and $K_n$ = $K_{nI}$.
We may now use Eqs.(5) and (6) to extract estimates of \ted\ from data for both $1/^{17}T_{1c}$ and $1/^{63}T_{1c}$.  
However, it is necessary to model \chpD\ in order to make systematic estimates of the $K_{nP}$.  
For this purpose we follow Aeppli, \textit{et al.} \cite{AeppliMason}, taking a squared 
Lorentzian form of unit amplitude $\chi^{\prime\prime}_P(\vec{q},\omega)/\chi^{\prime\prime}_{P}(peak)$ = $q_w^4/(q_w^2 + q_x^2 + q_y^2)^2$, from 
which the $K_{nP}$'s follow via Eq.(\ref{Kn}).  Since $^{17}T_{1c}$ (Eq.(\ref{17T1e2})) varies rapidly with $K_{1P}$, while $^{63}T_{1c}$ (Eq.(5)) is more weakly dependent on the $K_{nP}$'s, the width parameter $q_w$ may be varied with temperature to bring the $\tau_{eP}$'s into coincidence. Results of this procedure for LSCO are presented in Fig.1(b), with the corresponding values of $K_{nP}$ shown in the inset.

The squared Lorentzian form for \chpD\ centered on ($\pi,\pi$) gives a satisfactory account of the data,  
where we have taken, e.g., $K_{1P}(T)$ = -$0.81exp[-(T-50)/600]$.  The width parameter varies between $q_w\sim$ 0.6 and $\sim 1.3$ (units of 
$a^{-1}$) for $50K < T < 300K$.  $q_w$ is therefore similar to the displacement of the incommensurate peaks ($\sim 0.77$) in LSCO.

Next, we consider the case of YBCO6.5. While LSCO has only a weak pseudogap, YBCO6.5 has stood from the earliest days as a classic pseudogap system \cite{WWRW,Alloul,GotoYas}.  There now exists for YBCO6.5 a fairly complete, quantitative INS data set, discussed by the authors in terms of dynamical stripes \cite{Stock}.  Low-frequency data for \chpIa\ exhibit clear-cut $\omega/T$ scaling, yielding the horizontal solid line in Fig.\,2(a) for \tei.  The INS data show a 70/30 division between the populations of the two possible stripe domains and have a very nearly constant width parameter up to room temperature \cite{Stock}.  The YBCO6.5 data differ from LSCO in that the (INS) values of $K_{nI}$ lead, through Eq.(1) and (2) with measured HF constants \cite{HFConst}, to the widely disparate dash-dot curves for \ted\ in Fig.\,2(a).  Such a discrepancy in the region below $T_c\sim$ 62K suggests a sharp difference in the peak widths for \chpIa\ between the INS and NMR samples.  Indeed, broadening the \chpIa\ peaks by a factor $\sim$ 2.5 leads to unification of the \tef\ curves at T $<$ 62K, as shown by the blue square and red circle points in Fig.\,2(a).  This surprising broadening effect represents the difference between oriented powder samples used for the $T_1$ measurements \cite{Taki} and single crystals used for the INS studies, most likely due to different degrees of chain oxygen ordering.  Calculations of the coefficients $K_{nIa,b}$ made for Fig.\,2(a) used a form factor with discommensurations only along the $a$ axis based on the form given by Stock et al. \cite{Stock}.  The resulting temperature-independent correlation coefficients are $K_{1Ia}=-0.83$; $K_{1Ib}=-0.85$; $K_{2I}=0.74$; $K_{3Ia}=0.57$; $K_{3Ib}=0.61$.

\begin{figure}[htbp] 
\center
\includegraphics[bb=0.4in 0in 2.9in 5in, scale = 0.9]{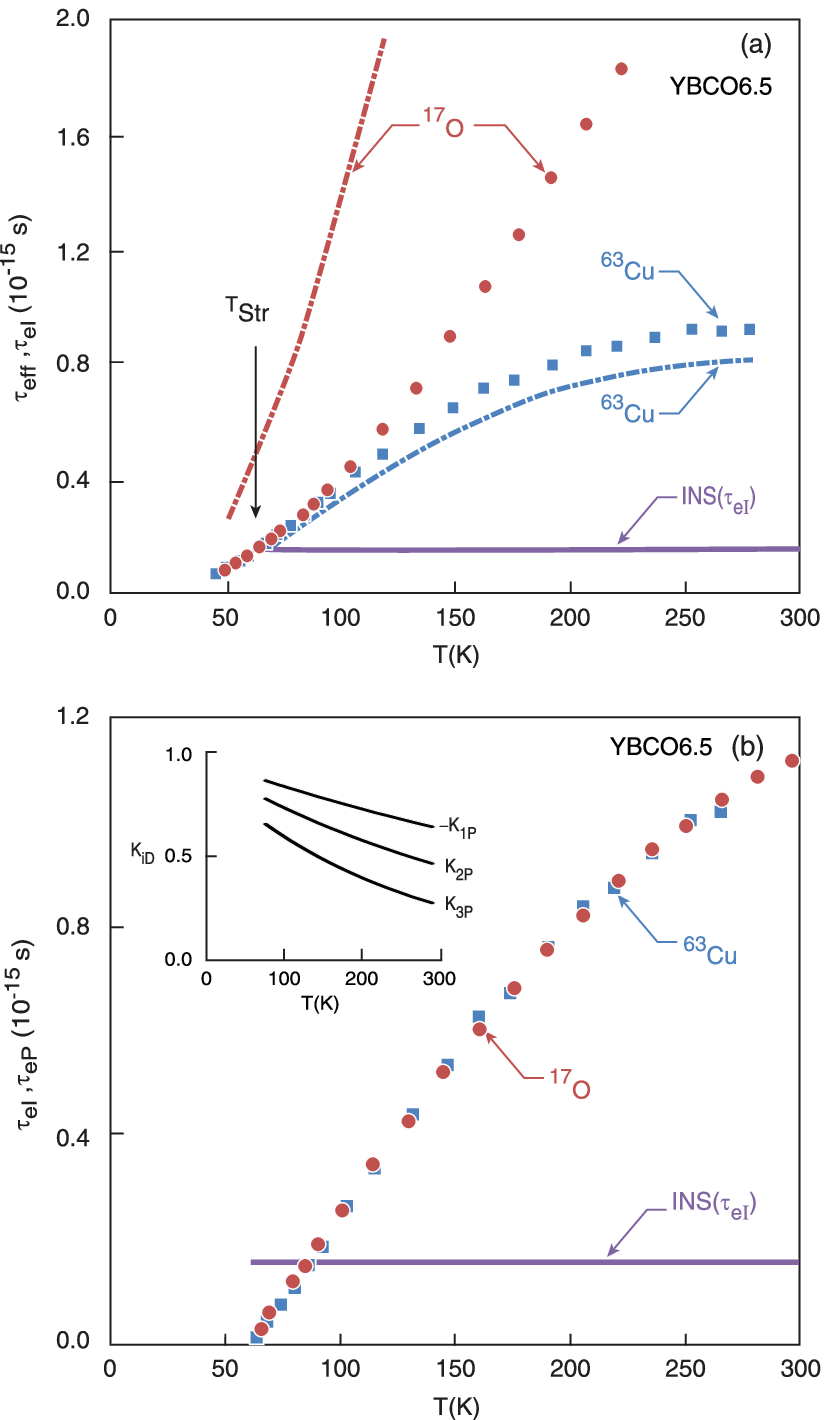}
\caption{ Plots of \tef, \tei, and \ted\ are presented for YBCO6.5 similar to the LSCO case in Fig.1. (a) Values of $\tau_{eI}(T)$ for YBCO6.5, calculated with Eq.(4) using INS data from Ref.\,7, are shown as a solid line that obeys $\omega/T$ scaling.  Values of \tef\ obtained with Eq.(1) for the \cu\ and with Eq.(2) for the \ox\ from $T_1$ data [21] using values of $K_{nI}$ are plotted as dash-dot lines. 
The disparity between the dash-dot lines for \cu\ and \ox\ is attributed to a disparity in incommensurate peak widths between NMR and INS samples and is corrected using adjusted peak widths (see text), leading to the curves showing filled squares (\cu ) and circles (\ox ).
At T = 62K the latter results show good mutual correspondence as well as agreement with \tei.
\,(b) Values of \ted\ obtained with Eq.(5) and (6) using $T_1$ data from Ref.\cite{Taki} are plotted against temperature. Calculation of the $K_{nP}$'s is described in the text.  A solid line representing the behavior of \tei\ from INS data (see part (a)) is replotted here for comparison. The $K_{nP}$'s with their fitted temperature dependences are shown in the inset. 
} 
\end{figure}

Values of \tef\ deduced from Eqs.(1) and (2) agree very nicely with \tei\ data (solid line) in Fig.\,2(a) at $T_c\sim$ 62K, 
again with no adjustable parameters.  
As with LSCO, the \tef\ curves show a sharp increase over \tei\ and a strong divergence from one another at T $>$ 62K.  To find consistent values of \ted\ for YBCO6.5,  we again model \chpD\ using the squared Lorentzian form as for LSCO with the same exponential form for
$K_{1P}(T)$.  The result (Fig.2(b)) is quite successful. Curves for the $K_{nP}$ are shown in the inset. In this case $K_{1P}(T)$
= -$0.87 exp[-(T-62)/725]$ decays a bit more slowly and begins with a somewhat narrower peak ($q_w\sim$ 0.44 at T = 62K).
The incommensurability $\sim 0.38$, however, is less than $q_w$, so that the progressively broadening peak of \chpD\ will form something of an
elevated baseline for the incommensurate peaks.  Such a background will be difficult to detect with unpolarized neutrons.
 
The \ted\ curves in Fig.\,1(b) and 2(b) are qualitatively similar, with \ted\ vanishing nearly linearly as T declines toward $T_{Str}$, while bending over towards room temperature.  At the latter point, the new term contributes far more to $1/T_1$ than the incommensurate 'stripe' fluctuations.  Values of \tei, which are considerably larger for LSCO than for YBCO6.5, obey $\omega/T$ scaling and also do not display a spin gap until the materials become bulk superconductors at $T_c$. The experimental conclusion is therefore clear - the much-celebrated magnetic pseudogaps in these systems are gaps in the new term deduced from $T_1$ data, which accounts for more spectral weight than the incommensurate spin fluctuations at NMR frequencies. 
Moreover, given the strong evidence that the stripe (incommensurate) fluctuations which dominate low- and medium energy neutron measurements compete with superconductivity, it is the pseudogap terms that are much more likely to form a pair binding texture.

Some time ago, in the first quantitative test of the magnetic fluctuation-dissipation 
theorem a joint analysis of NMR-INS data on LSCO was presented \cite{WSC}. That work was 
only a partial success because of its rather simple treatment of the $T_1$ process. At the time, 
a two-band model was called for, but there is no longer any clear motivation for such a 
model \cite{BPT}. However, NMR shift analyses have been put forth recently giving evidence for a 
'two-component' shift structure \cite{Haase}.  We suggest that the two-part structure for nuclear relaxation 
described in the present work could, via the Kramers-Kronig relation, form the basis 
in principle for a two-component NMR shift. In practice, there are no \chipp\ data near 
$\vec{q}$ = $\vec{0}$ to provide a quantitative basis for a shift estimate. However, the proposed 
NMR shift structure is regarded as a natural extension of the present two-component model for \chipp. 
We emphasize that the latter model does not imply two independent bands of charge carriers.

In closing, we comment on the prospects for resolving the new fluctuation term \chpD\ with INS methods.
The squared Lorentzian model form factor used here yields estimates for \chpD\ that are generally much broader and flatter than the incommensurate peaks reported to date. For LSCO, such a model yields an amplitude $\sim$10\% of \chpIa\ for a scan through adjacent peaks.  Such a result is compatible with experimental spectra for $50K\leq T\leq 100K$\cite{AeppliMason}.  Thus, resolving \chpD\ will require polarized neutrons with a far higher signal-to-noise ratio than what has been reported to date. Recent improvements in neutron scattering technique may render this feasible.  However, we conclude that the \chpD\ term, whatever its precise shape and behavior may be, must exist, rendering \chipp\ consistent with pseudogap studies using other probes. We suggest this inference to extend to other cuprates as well, offering a clear and broadly consistent picture of the pseudogap effect throughout this family of superconductors.

We thank B. S. Shastry and J. Zaanen for highly informative conversations. This manuscript has been authored by UT-Battelle, LLC, under Contract No. DE-AC05-00OR22725 with the U.S. Department of Energy. 


\begin{thebibliography}{AeppliMason}
\bibitem {Error} We suggest error limits of $\pm 30\%$ for INS-derived \tef\ values for both LSCO and YBCO6.5.
\bibitem {HFConst} The HF constants ($A_{ab},B,C_a,C_b$) used are (19, 76, 108, 64) for LSCO [20] and (31, 79, 145, 89) for YBCO6.5 [21], respectively, in units of kG/spin. $\gamma_{17}/\pi$ = 577.19 Hz/G; $\gamma_{63}/\pi$ = 1128.5 Hz/G.
\end{thebibliography}
\end{document}